\documentclass[pre,aps,preprint,eqsecnum]{revtex4}
\usepackage{epsfig}
\usepackage{bm}
\newcommand{\B}[1]{{\bm{#1}}}
\newcommand{\C}[1]{{\mathcal{#1}}}
\usepackage[latin1]{inputenc}
\newcommand{\beq}{\begin{equation}}
\newcommand{\eeq}{\end{equation}}
\newcommand{\bea}{\begin{eqnarray}}
\newcommand{\eea}{\end{eqnarray}}
\begin{document}

\title{Free-Boundary Dynamics in Elasto-plastic Amorphous Solids: \\
The Circular Hole Problem}
\author{Eran Bouchbinder$^*$, J. S. Langer$^\dag$, Ting-Shek Lo$^*$ and Itamar Procaccia$^*$ }
\address{$^*\!$Dept. of Chemical Physics, The Weizmann Institute of Science, Rehovot 76100, Israel,
\\$^\dag$Dept. of Physics, University of California Santa Barbara}


\begin{abstract}
We develop an athermal shear-transformation-zone (STZ) theory of
plastic deformation in spatially inhomogeneous, amorphous solids.
Our ultimate goal is to describe the dynamics of the boundaries
of voids or cracks in such systems when they are subjected to
remote, time-dependent tractions. The theory is illustrated here
for the case of a circular hole in an infinite two-dimensional
plate, a highly symmetric situation that allows us to solve much
of the problem analytically. In spite of its special symmetry,
this example contains many general features of systems in which
stress is concentrated near free boundaries and deforms them
irreversibly. We depart from conventional treatments of such problems
in two ways.  First, the STZ analysis allows us to keep track of
spatially heterogeneous, internal state variables such as the effective
disorder temperature, which determines plastic response to subsequent
loading. Second, we subject the system to stress pulses of finite
duration, and therefore are able to observe elasto-plastic response
during both loading and unloading. We compute the final deformations
and residual stresses produced by these stress pulses.
Looking toward more general applications of these results,
we examine the possibility of constructing a boundary-layer theory that
might be useful in less symmetric situations.
\end{abstract}
\maketitle

\section{Introduction}

The aim of this paper is to develop a theoretical description of
the way in which the boundaries of voids in
amorphous solids move under the influence of external loads.
An obvious example is dynamic fracture.  The evolution of
cracks is traditionally described by linear elasticity.
However, elasticity by itself does not provide a dynamical theory
of crack motion but, instead, uses criteria such as energy balance to
provide bounds on the velocities of straight cracks \cite{Freund}.
This approach was challenged by one of us in \cite{00Lan}. More
recently, starting with closely related ideas, it was shown in
\cite{06BPP} that taking into account the dynamic degrees of freedom
near a moving crack tip resolves various inconsistencies in
traditional theories of linear elasticity fracture mechanics.
Nevertheless, the theoretical approach of \cite{06BPP} is far
from complete; it is still not an accurate and internally
self-consistent description of free-boundary motion.

We do not deal with the actual dynamics of cracks in this
paper. Rather, we apply the shear-transformation-zone (STZ)
theory developed in two recent papers
\cite{06BLPa, 06BLPb} to the dynamics of a circular hole in an
infinite medium. We have chosen this highly symmetric example in
order to simplify the tensorial character of the theory, and to help
in developing analytic methods that may remain applicable
in less symmetric situations. As we shall see, the solutions of the circle problem
provide a wealth of information about the ways in which stress
concentrations near moving free boundaries induce plastic
deformation.

The STZ approach that we employ here is based on recent developments in
the theory of dynamic elasto-plastic deformation of amorphous
materials at low temperatures \cite{06BLPa}. The
theory was applied to a homogenous situation in \cite{06BLPa} and the results
were compared to recent numerical simulations in \cite{06BLPb}. The STZ theory,
originally proposed in \cite{Early_STZ,03LP}, deviates from conventional
approaches \cite{Lubliner} by focusing on the nature of the
microscopic mechanisms for plastic deformation in amorphous media.
These mechanisms are incorporated into the macroscopic description
by internal state fields. An important observation is that the
plastic strain itself cannot be one of these fields. For that to
be true, the material would somehow have to encode information
about its entire history of deformation starting from some
reference state.  This is not possible.  More realistically,
the memory of prior deformations, and the rate at which that memory
is lost, must be encoded in the internal
state variables and their equations of motion.  The identification
of the physically appropriate microscopic variables is based here on the notion that STZ's
are sparsely distributed, localized clusters of molecules
that are especially susceptible to rearrangement in response to
applied stresses. To take the STZ's and their role in the dynamics into account,
one introduces a scalar field that describes
the density of STZ's and a tensorial one that describes their
orientation. Equations of motion for both the plastic strain rate
and these internal state fields are obtained by assuming that
STZ's change orientation (deform by finite amounts) at rates
that depend on the stress. In addition, STZ's are created and
annihilated at rates proportional to the rate at which the energy
of plastic deformation is dissipated \cite{03LP}.

An earlier STZ analysis of the circle problem \cite{99LL}
was based on a ``quasilinear" version of the theory that
used a conveniently truncated form of the transition rates
\cite{Early_STZ,03LP}. This relatively simple theory captured linear
visco-elasticity at small stresses, finite visco-plasticity at
intermediate stresses, and unbounded plastic deformation at large
stresses as a result of an exchange of dynamic stability between
jammed (non-deforming) and unjammed (deforming) states of the
system. Some memory effects were also described successfully.
The quasilinear theory, however, had serious limitations; and therefore
we base the present analysis on the recent version of STZ theory
\cite{06BLPa} that starts with more physically realistic assumptions.
In \cite{06BLPa}, it was argued that the defining
feature of a system at low temperatures is the constraint that,
because thermal activation of STZ transitions is negligible or
nonexistent, each molecular rearrangement occurs in response to an
external driving force. No motion occurs in the absence of such a
force, and no rearrangement moves in opposition to the direction of
that force. In addition, this version of STZ theory includes the
possibility that STZ's occur in many different types and with a range
of different transition thresholds.

We list here some important respects
in which the present analysis differs from the previous work on
the circular hole problem in \cite{99LL} and elsewhere \cite{Refs}:
\begin{enumerate}

\item We employ here the fully nonlinear, athermal STZ theory
described in \cite{06BLPa}.

\item The material time derivatives are expressed in a
proper Eulerian formulation; and the theory is consistent with
all relevant conservation laws and symmetries.

\item The calculation described here takes into account the
dynamics of the density of STZ's {\it via} an effective-temperature
formulation, a feature that is shown in \cite{06BLPa, 06BLPb}
to be essential for self-consistency in these theories.

\item The focus here is on the transient growth regime that is
relevant to dynamic situations such as fracture.  Instead of
considering only constant loads, we look at the elasto-plastic
response to a stress pulse, and study the behavior during
both loading and subsequent unloading.

\end{enumerate}

The organization of this paper is as follows: In Sec. \ref{EqBC} we write
the equations of motion and the boundary conditions for a circular hole in an
infinite sheet of material subjected to a uniform radial stress at infinity.
In Sec. \ref{STZ_equations}, we review the elements of the STZ theory
that are needed for this problem and specialize to cases of interest here.
Sec. \ref{loading} contains discussions
of the specific situations to be considered in more detail, starting with
the two different loading schemes to be used, and then the simplifying
assumption of elastic incompressibility.
Our numerical results for both constant applied stress and two different
stress pulses are presented in Sec. \ref{results}. Sec. \ref{BLsect} is
devoted to the development of a boundary layer approximation and its
comparison with the numerical results. Finally, in Sect. \ref{summary},
we offer a summary and further discussion.  In Appendix \ref{cavitation},
we show how to compute the threshold for unbounded growth of the circular hole
under constant remote loading.

\section{Equations and Boundary Conditions}
\label{EqBC}

\subsection{Strain, stress and rate of deformation}

We start by writing the full set of equations for a general
two-dimensional elasto-plastic material. We define the total
rate-of-deformation tensor
\begin{equation}
D^{\rm tot}_{ij} \equiv \frac{1}{2}\Big(\frac{\partial v_i}{\partial
x_j} + \frac{\partial v_j}{\partial  x_i}\Big)  \ , \label{D_tot}
\end{equation}
where $\B v(\B r, t)$ is the material velocity at the location $\B
r$ at time $t$. This type of Eulerian formulation has the
advantage that it disposes of any reference state, allowing free
discussion of small or large deformations. The price that we pay is that,
in inhomogeneous situations, we need to employ the full material
derivative for a tensor $\B A$,
\begin{equation}
\frac{{\cal D} \B A}{{\cal D} t} = \frac{\partial \B A}{\partial t}
+ \B v\cdot \B\nabla \B A +\B A \cdot \B \omega - \B \omega\cdot \B
A \ , \label{material}
\end{equation}
where $ \B\omega$ is the spin tensor
\begin{equation}
\omega_{ij} \equiv  \frac{1}{2}\left(\frac{\partial v_i}{\partial
 x_j} -\frac{\partial v_j}{\partial x_i}\right)  \ . \label{omega}
\end{equation}
For a scalar or vector quantity $\B A$ the commutator with the spin
tensor vanishes identically.

Plasticity is introduced by assuming that the total
rate-of-deformation tensor can be written as a sum of
elastic and plastic contributions
\begin{equation}
D^{\rm tot}_{ij}  = \frac{{\cal D} \epsilon^{\rm el}_{ij}}{{\cal D}
t} +D^{\rm pl}_{ij} \label{el_pl} \ . \label{separate}
\end{equation}
The linear elastic strain tensor $\epsilon^{el}_{ij}$ is related to
the stress tensor whose general form is
\begin{equation}
\sigma_{ij} = -p\delta_{ij} + s_{ij} \ , \quad p=-\case{1}{2}
\sigma_{kk} \ . \label{sig}
\end{equation}
The relation is
\begin{equation}
\epsilon^{\rm el}_{ij} = -\frac{p}{2K}\delta_{ij} +
\frac{s_{ij}}{2\mu} \ , \label{linear}
\end{equation}
where $K$ and $\mu$ are the two dimensional bulk  and
shear moduli respectively.  $s_{ij}$ is the deviatoric stress tensor,
and $p$ is the pressure. The equations of acceleration and continuity are
\begin{eqnarray}
\rho \frac{{\cal D} \B v }{{\cal D} t} &=& \B \nabla\!\cdot\!\B
\sigma = -\B \nabla p+ \B \nabla\!\cdot\! \B s \label{eqmot1} \ , \\
\quad \frac{{\cal D} \rho}{{\cal D} t}  &=& -\rho \B \nabla\!
\cdot\! \B v \ . \label{eqmot2}
\end{eqnarray}

Consider now an infinite two-dimensional amorphous material with
a circular hole centered at the origin. The hole has
radius $R$; and the system is loaded at infinity by a radial,
uniform and possibly time-dependent stress $\sigma^{\infty}(t)$.
For this configuration, the field of interest is the radial
velocity  $v_r(r,t)$, denoted here simply by $v(r,t)$.
The non-vanishing components of $D^{tot}_{ij}$ in Eq. (\ref{D_tot})
are
\begin{equation}
D^{tot}_{rr} = \frac{\partial v}{\partial
 r} \ , \quad  D^{tot}_{\theta \theta} = \frac{v}{
 r} \label{D_components} \ .
\end{equation}
Note that $D^{tot}_{ij}$ satisfies the compatibility relation
\begin{equation}
D^{tot}_{rr} = \frac{\partial}{\partial
 r} \left(r D^{tot}_{\theta \theta}\right) \label{compatibility} \ .
\end{equation}
The azimuthal velocity vanishes identically. The material time
derivative in Eq.(\ref{material}) becomes
\begin{equation}
\frac{\cal D}{{\cal D} t} \equiv \frac{\partial}{\partial
 t} + v \frac{\partial}{\partial
 r}\label{Mat_Der} \ .
\end{equation}

We now turn to the equations of motion (\ref{eqmot1}). The symmetry
of the problem implies that only the $\sigma_{rr}$ and
$\sigma_{\theta \theta}$ components of the stress tensor are
non-vanishing.  The tensor $s_{ij}$ is diagonal, and
thus we define the deviatoric stress $s$ and the pressure $p$ to be
\begin{equation}
p \equiv -\frac{1}{2}(\sigma_{\theta \theta}+\sigma_{rr}), \quad s
\equiv s_{\theta \theta}=-s_{rr}= \frac{1}{2}(\sigma_{\theta
\theta}-\sigma_{rr})\label{Pr_Devia} \ .
\end{equation}
Next we assume that the inertial term on the left-hand side
of Eq.(\ref{eqmot1}) is negligible and that density variations
are also negligible; thus Eq.(\ref{eqmot2}) is satisfied automatically,
and Eqs.(\ref{eqmot1}) reduce to a single force balance equation
of the form \cite{86LL}
\begin{equation}
\frac{\partial \sigma_{rr}}{\partial r} +
\frac{\sigma_{rr}-\sigma_{\theta \theta}}{r} = 0 \label{Force_Bal_1}
\ ,
\end{equation}
which, with Eqs.(\ref{Pr_Devia}), becomes
\begin{equation}
\frac{\partial p}{\partial r}
=-\frac{1}{r^2}\frac{\partial}{\partial
 r} (r^2s) \label{Force_Bal_2}
\ .
\end{equation}
The hole edge is a free boundary, {\it i.e.} it is traction-free;
and therefore the boundary conditions are:
\begin{eqnarray}
\sigma_{rr}(R,t)=-p(R,t)-s(R,t)=0 \nonumber\\
p \to -\sigma^{\infty}(t), \quad s \to 0 \quad \hbox{as} \quad r \to
\infty \label{BC} \ ,
\end{eqnarray}
where $\sigma^{\infty}(t)$ is assumed to vary slowly enough
that the omission of the inertial terms in Eq. (\ref{eqmot1}) is
still justified.

\subsection{Introducing plasticity}

Plasticity is introduced as in Eq.(\ref{separate}). According to
Eq.(\ref{linear}), the components $\epsilon^{el}_{ij}$ of the strain
tensor satisfy
\begin{eqnarray}
\epsilon^{el}_{rr}&=&-\frac{p}{2K}-\frac{s}{2\mu}\ , \nonumber\\
\epsilon^{el}_{\theta
\theta}&=&-\frac{p}{2K}+\frac{s}{2\mu}\label{Hooke} \ .
\end{eqnarray}
The two-dimensional bulk modulus $K$ is given by
\begin{equation}
K = \frac{\mu (1+\nu^*)}{1-\nu^*} \ , \label{bulk_mod}
\end{equation}
where $\nu^*$ is the two-dimensional Poisson ratio. The plastic
rate-of-deformation tensor $D^{pl}_{ij}$ is assumed to be traceless
(incompressible plasticity), which in circular symmetry means
\begin{equation}
D^{pl}_{\theta \theta}=-D^{pl}_{rr} \equiv D^{pl} \ . \label{D}
\end{equation}
The constitutive behavior is introduced by writing
$D^{pl}$ as a function of the deviatoric stress $s$ and the
internal state fields. The material velocity at the edge of the hole
$v(R,t)$ is just the rate of change of the hole radius
\begin{equation}
\dot{R}(t)=v(R,t)  \label{edge_velocity}
\end{equation}
and therefore, using the second equation in (\ref{D_components}),
we see that the hole evolves according to
\begin{equation}
\frac{\dot{R}}{R}= D^{tot}_{\theta \theta}(R,t) \label{evolution} \
.
\end{equation}
Once $D^{tot}_{\theta \theta}(R,t)$ is specified, the last
relation becomes an equation of motion for the boundary of the hole.

\subsection{The velocity field}

In order to explore the analytic structure of the problem, we
use Eqs. (\ref{el_pl}), (\ref{D_components}), (\ref{Hooke}) and
(\ref{D}) to obtain
\begin{eqnarray}
\frac{v}{r}+\frac{\partial v}{\partial r} &=&
-\frac{1}{K}\frac{{\cal D}p}{{\cal D}
t}= -\frac{1}{K}\left[\dot{p}+v \frac{\partial p}{\partial r}\right]\ , \label{kinematic1} \\
\frac{v}{r}-\frac{\partial v}{\partial r}&=&
\frac{1}{\mu}\frac{{\cal D}s}{{\cal D} t} +2D^{pl}=
\frac{1}{\mu}\left[\dot{s}+v \frac{\partial s}{\partial r}\right]
+2D^{pl} \label{kinematic2} \ ,
\end{eqnarray}
where dots denote partial time derivatives. It is useful next to eliminate
$\dot{s}(r,t)$ and $\dot{p}(r,t)$ from these equations, and thus to obtain a
differential equation for the velocity field $v(r,t)$ that does not depend
explicitly on those time derivatives.  For that purpose, we operate with
$-(\mu/K\,r^2)\,(\partial/\partial r)\,[r^2(\cdot)]$
on Eq.(\ref{kinematic2}), differentiate Eq.(\ref{kinematic1})
with respect to $r$, add the results, and then use the partial time
derivative of Eq.(\ref{Force_Bal_2}) to obtain
\begin{eqnarray}
&&\left(1+\frac{\mu}{K}\right)\left(\frac{\partial^2 v}{\partial
r^2}+\frac{1}{r}\frac{\partial v}{\partial
r}-\frac{v}{r^2}\right)= \nonumber\\
&&-\frac{1}{K}\left[\frac{\partial v}{\partial r}\frac{\partial
p}{\partial r}+v\frac{\partial^2 p}{\partial
r^2}+\frac{1}{r^2}\frac{\partial}{\partial
 r} \left(r^2 v \frac{\partial s}{\partial
 r}\right)
\right] \nonumber\\
&&-\frac{\mu}{K} \left(\frac{4 D^{pl}}{r}+s\frac{\partial
D^{pl}}{\partial r} \right) \label{v_1} \ .
\end{eqnarray}
This is a linear second order differential equation for $v$ with
coefficients that are functions of $s(r,t)$ and $p(r,t)$. This equation
can be further simplified by defining $\tilde{v}\!\equiv\! v/r$ and
operating with $\partial_r [v (\cdot)]$
on Eq.(\ref{Force_Bal_2}). After simple manipulations we obtain
\begin{eqnarray}
\label{v_2} \left( 1+\frac{\mu}{K} \right)\left( r \frac{\partial^2
\tilde{v}}{\partial r^2}+ 3 \frac{\partial \tilde{v}}{\partial r}
\right) = \nonumber\\
 \frac{2 s }{K} \frac{\partial
\tilde{v}}{\partial r} - \frac{\mu}{K} \left(\frac{4
D^{pl}}{r}+s\frac{\partial D^{pl}}{\partial r} \right) \ .
\end{eqnarray}
This equation requires two boundary conditions.
To obtain these, we note that the first boundary condition of
Eq.(\ref{BC}) implies
\begin{equation}
\frac{{\cal D} [p(R,t)+s(R,t)]}{{\cal D} t} = 0 \ . \label{der_BC}
\end{equation}
Then, using Eqs.(\ref{kinematic1}) and (\ref{kinematic2}) and simple
manipulations, we obtain
\begin{equation}
\label{BC_v1} \frac{\partial \tilde{v}(R,t)}{\partial r} = -\frac{
2K \tilde{v}(R) - 2\mu D^{pl}(R,t)}{(K + \mu)R}.
\end{equation}

To complete the derivation we
need an expression for $\tilde{v}(R)$ in terms of $s$ and
$D^{pl}$. To that aim we integrate Eq.(\ref{Force_Bal_2})
from $R(t)$ to $\infty$ and use the first
boundary condition of Eq.(\ref{BC}) to obtain the
integral relation:
\begin{equation}
\sigma^{\infty}(t)=2\int_{R(t)}^{\infty} \frac{s(r,t)}{r} dr \ .
\label{integral}
\end{equation}
Taking the time derivative of Eq.(\ref{integral}), we obtain
\begin{equation}
\int_{R(t)}^{\infty} \frac{{\cal D} s(r,t)}{{\cal D} t}
\frac{dr}{r}=\frac{s(R,t) \dot{R}}{R} +
\frac{\dot{\sigma}^{\infty}(t)}{2} \ . \label{diff_integral}
\end{equation}
In order to use the last result we operate with
$2\int_R^{\infty}(\cdot) (dr/r) $
on Eq. (\ref{kinematic2}) to obtain
\begin{equation}
\tilde{v}(R)-\tilde{v}(\infty)=\frac{1}{\mu} \int_{R(t)}^{\infty}
\frac{{\cal D} s(r,t)}{{\cal D} t} \frac{dr}{r}+
2\int_{R(t)}^{\infty} \frac{D^{pl}}{r}dr \ .
\label{operate_result}
\end{equation}

To evaluate $\tilde{v}(\infty)$, note that, for
$r\gg R$, the stress field is
purely elastic and therefore the solution of Eq.(\ref{Force_Bal_2}) is
$p=-\sigma^{\infty}$ and $s \sim r^{-2}$. Then, the large-$r$
behavior of Eqs.(\ref{kinematic1})-(\ref{kinematic2}) predicts that
\begin{equation}
\label{BC_v2} \tilde{v}(\infty)=\frac{\dot{\sigma}^{\infty}}{2K} \,.
\end{equation}
Substituting Eqs.(\ref{diff_integral}) and
(\ref{BC_v2}) in Eq.(\ref{operate_result}) we find
\begin{equation}
\frac{\dot{R}}{R} = \tilde{v}(R) =  \frac{\displaystyle
2\int_{R}^{\infty} \frac{D^{pl}(r,t)}{r}dr
+\dot{\sigma}^{\infty}(t)\left(\frac{1}{2\mu}+\frac{1}{2K}\right)}
{\displaystyle 1-\frac{s(R,t)}{\mu}} \ . \label{BC_v3}
\end{equation}
This equation plays a key role throughout the rest of this paper.
Note that it already incorporates both the boundary conditions of Eq. (\ref{BC}) and that,
together with Eq. (\ref{BC_v1}), it provides the
boundary conditions that we need to solve Eq. (\ref{v_2}) for the velocity
field $v(r,t)$.
\subsection{Putting it all together}

In order to put the equations in their final forms, up to specifying
$D^{pl}$, we define
\begin{equation}
W(r,t) \equiv \frac{\partial \tilde{v}(r,t)}{\partial r} \ .
\label{W}
\end{equation}
Then Eq.(\ref{v_2}) becomes
\begin{eqnarray}
\label{v_3} &&\left( 1+\frac{\mu}{K} \right)\left( r \frac{\partial
W(r,t)}{\partial r}+ 3 W(r,t)
\right) = \nonumber\\
 &&\frac{2 s(r,t) }{K} W(r,t) - \frac{\mu}{K} \left(\frac{4
D^{pl}(r,t)}{r}+s(r,t)\frac{\partial D^{pl}(r,t)}{\partial r}
\right) \ , \nonumber\\
\end{eqnarray}
which is to be solved using the boundary conditions given in Eqs.(\ref{BC_v1}) and
(\ref{BC_v3}).  The former becomes
\begin{equation}
 W(R,t) = -\frac{\displaystyle 2K
\frac{\dot{R}}{R} - 2\mu D^{pl}(R,t)}{\displaystyle (K + \mu)R} \ .
\end{equation}
In terms of $W(r,t)$, the equation of motion for the deviatoric stress
$s(r,t)$, Eq.(\ref{kinematic2}), is
\begin{equation}
-rW(r,t)= \frac{1}{\mu}\left[\dot{s}(r,t)+v(r,t) \frac{\partial
s(r,t)}{\partial r}\right] +2D^{pl}(r,t) \label{s_eq} \ .
\nonumber\\
\end{equation}
The pressure $p(r,t)$ can be calculated using Eq.
(\ref{Force_Bal_2}). According to Eq. (\ref{W}), the velocity field
$v(r,t)$ is given by
\begin{equation}
v(r,t)=\frac{\dot{R} r}{R} + r\int_R^r W(r',t)dr'  \ .  \label{v_W}
\end{equation}
The initial conditions are the linear elastic
fields determined by the load applied at $t\!=\!0$
\begin{eqnarray}
p(r,t=0)&=&-\sigma^{\infty}(0)\ ,  \nonumber\\
s(r,t=0)&=& \sigma^{\infty}(0) \frac{R^2(0)}{r^2} \label{initial} \ .
\end{eqnarray}

\section{Plastic Deformation}
\label{STZ_equations}

To complete this theory, we need to choose a specific plastic
rate-of-deformation function $D^{pl}(s,...)$ and equations
of motion for the internal fields (denoted by the dots).
As stated in the Introduction, we use the
STZ theory described in reference \cite{06BLPa}.
Here, $D^{pl}(s,\Lambda,m)$
depends on the deviatoric stress $s$, on the normalized STZ
density $\Lambda$, and on a field $m$ that describes the local
average STZ orientation. Although ordinary thermal fluctuations are
absent at the low temperatures considered here, the concept of an
effective disorder temperature remains essential \cite{06BLPa,
06BLPb}. The effective temperature $T_{eff}$
characterizes the state of configurational disorder in the system.
It equilibrates to the ambient temperature $T$ at high
$T$ (relative to the glass temperature $T_g$),
but may fall out of equilibrium at low $T$ where disorder is
generated by the atomic-scale, configurational rearrangements that
accompany mechanical deformations.  We denote $T_{eff}=
(E_{STZ}/k_B)\,\chi$, where $E_{STZ}$ is a characteristic STZ
formation energy;  then the STZ density is proportional to the
Boltzmann factor $\exp\,(-1/\chi)$. Most importantly, the time
variation of the STZ density is slaved to the dynamics of
$\chi$.

The set of equations that describes plasticity in our theory
\cite{06BLPa} is:
\begin{eqnarray}
&& D^{pl}(\tilde
s,m,\Lambda)={\epsilon_0\over\tau_0}\,\Lambda\,q(\tilde s,m)\label{Dplav3}\ , \label{Dplqsm}\\
&& \frac{{\cal D} m}{{\cal D} t} = {2\over \tau_0}\,q(\tilde
s,m)\,\Bigl(1 -
{m\,\tilde s\over \Lambda}\,e^{-1/\chi}\Bigr)\label{dotm3}\ , \\
&&  \frac{{\cal D} \Lambda}{{\cal D} t} = {2\over\tau_0}\,\tilde
s\,q(\tilde
s,m)\,\bigl(e^{-1/\chi} - \Lambda\bigr)\label{dotLambda3}\ , \\
&&  \frac{{\cal D} \chi}{{\cal D} t} = {2\,\epsilon_0\over
c_0\,\tau_0}\,\Lambda\,\tilde s\,q(\tilde s,m)\,(\chi_{\infty}-\chi)
\label{dotchi3} \ ,
\end{eqnarray}
where
\begin{equation}
q(\tilde s,m) \equiv \C C(\tilde s)\,\left({\tilde s\over |\tilde
s|}-m\right) \ , \label{defqsm}
\end{equation}
and $\tilde s=s/s_y$, with $s_y$ being the dynamic yield stress
\cite{06BLPa}. Here $\tau_0$ is an elementary time scale,
$\epsilon_0$ is a dimensionless constant of order unity, $c_0$ is a specific heat
expressed in units of $k_B$ per atom and thus is of order
unity, and $\chi_{\infty}$ is the asymptotic value approached by
$\chi$ during continuous deformation \cite{06BLPa}.

Note that many of the features of these equations are model independent.
For example, Eqs. (\ref{Dplqsm}) and (\ref{defqsm}) identify $m$ as
a ``back stress" that governs an exchange
of dynamic stability at $|\tilde s|=1$ (i.e. when
$s=s_y$). For $|\tilde s|<1$, the stable steady-state solutions of
these equations occur at $|m|=1$ such that the system is jammed and
$D^{pl}(\tilde s,m,\Lambda)=0$. For $|\tilde s|>1$, on the other hand,
$m =1/\tilde s$ and there is a non-vanishing plastic flow
(see \cite{06BLPa} for more details).

Essentially all of the specific
material-dependent properties of this theory are contained in the
function ${\cal C}(\tilde s)$. In the STZ model described in
\cite{06BLPa}, ${\cal C}(\tilde s)$ is related to the rate $R(\tilde
s)$ at which STZ's transform between their two orientations:
\begin{equation}
\C C(\tilde s) \equiv \frac{\tau_0[R(\tilde s)+R(-\tilde s)]}{2} \ ,
\end{equation}
where $R(\tilde s)$ is given by
\begin{eqnarray}
R(\tilde s)&=&\frac{2}{\tau_0}\int_0^{\tilde{s}} (\tilde s-\tilde
s_\alpha)~P(\tilde s_\alpha;\zeta)~d\tilde s_\alpha\label{integral2}
\ , \\
P(\tilde s_\alpha;\zeta)&=& \frac{\zeta^{\zeta+1}}{\zeta!}~\tilde
s_\alpha^\zeta~e^{-\zeta \tilde s_\alpha} \label{pdf} \ .
\end{eqnarray}
Here, $R(\tilde s)$ is an integral over a distribution $P(\tilde
s_\alpha;\zeta)$ that reflects the fact that the STZ's can be of various
types with different activation thresholds $\tilde s_{\alpha}$.  The
parameter $\zeta$ that characterizes the
distribution is the only material parameter here; it controls the
width of the distribution. The mean value of the distribution is
$s_y$, the dynamic yield stress, that was shown in \cite{06BLPa}
to be the value of the deviatoric
stress at which the system undergoes its dynamic exchange of stability
from jammed to flowing states. For finite values of $\zeta$
there can be nonzero, sub-yield plastic deformation for $|\tilde s|<1$.
This behavior is well documented in the
literature \cite{Lubliner}. We note that for $\tilde s$ very small
or very large,
\begin{eqnarray}
R(\tilde s) &\sim& \tilde s^{\zeta+2} \quad\hbox{for}\quad \tilde s \to 0^+ \\
R(\tilde s) &\simeq& \tilde s-1 \quad\hbox{for}\quad \tilde s \gg 1
\ .
\end{eqnarray}

A basic assumption of the STZ theory is that the STZ's are sparsely
distributed and only weakly interacting with each other.
For this assumption to be valid, $\chi_{\infty}$ must small.
Indeed, in independent applications of the theory to actual
materials \cite{04Lan} or simulations of such materials
\cite{06BLPb, 06SKLF}, $\chi_{\infty}$ was found to be of order
$0.15$ or less. For such values of $\chi_{\infty}$, the density of STZ's,
$\Lambda \cong \exp\,(-1/\chi_{\infty})$, is of order $10^{-3}$, or very much
smaller. We then notice that $\Lambda$ appears as a rate-determining pre-factor on
the right-hand sides of Eqs. (\ref{Dplav3}) and (\ref{dotchi3}),
which govern the bulk system-wide variables $D^{pl}$ and $\chi$; but
$\Lambda$ does not appear in a similar way in Eqs. (\ref{dotm3}) or
(\ref{dotLambda3}), which pertain to the dynamics of  individual
STZ's.  It follows that the plastic strain rate and the effective
temperature respond much more slowly to changes in stress than do the
internal fields $m$ and $\Lambda$, and that the slow
dynamics of the effective temperature controls the observable
mechanical behavior of the system in most circumstances. We conclude
that, as long as the characteristic time scale of the external
loading is not significantly smaller than $\tau_0
\exp\,(1/\chi_{\infty})$, we can safely replace Eqs.(\ref{dotm3})
and (\ref{dotLambda3}) by their stationary solutions:
\begin{equation}
\label{m0}
m=m_0(\tilde s)=\cases{\tilde s/ |\tilde s| &if $|\tilde s|\le 1$\cr 1/\tilde s & if $|\tilde s| >1$}
\end{equation}
and
\begin{equation}
\Lambda = e^{-1/\chi} \ . \label{Lamchi}
\end{equation}
The approximation in Eq.(\ref{m0}) tells us
that $|m|=1$ for all $|\tilde s| < 1$, {\it i.e.} the system ``immediately"
becomes jammed at small stresses.  In fact, one can think of an experiment in which at time $t=0$
a virgin material with $m=0$ throughout is being used. Then in regions where $\tilde s$ is small,  $m$ will remain very small for a long time, since $\C C(\tilde s)$ is small. Notwithstanding,  the precise value
of $m$ in regions of small $\tilde s$ is not important for any of the results presented below. Therefore, we use Eqs. (\ref{m0}) and (\ref{Lamchi}) to reduce  Eqs. (\ref{Dplav3})-
(\ref{defqsm}) to
\begin{eqnarray}
&& D^{pl}(\tilde
s,m,\Lambda)={\epsilon_0\over\tau_0}\,e^{-1/\chi}\,q_0(\tilde s),~~~~q_0(\tilde s) \equiv \C C(\tilde s)\,\left({\tilde s\over |\tilde
s|}-m_0(\tilde s)\right)\label{STZ1}\ , \\
&&  \frac{{\cal D} \chi}{{\cal D} t} = {2\,\epsilon_0\over
c_0\,\tau_0}\,e^{-1/\chi}\,\tilde s\,q_0(\tilde
s)\,(\chi_{\infty}-\chi) \label{STZ3} \ .
\end{eqnarray}
Once having made these simplifications, we also may go immediately
to the limit $\zeta \to \infty$ and neglect any structure in the
function ${\cal C}(\tilde s)$  for stresses less than or of order
the yield stress. Without this approximation, our
equations of motion would have predicted sub-yield deformations
of magnitude proportional to
$\C C(\tilde s) \exp(-1/\chi_{\infty})$, which we assume to be
extremely small.  Such deformations will not be negligible, of course,
for soft and/or highly disordered systems, or for thermal systems at
temperatures high enough to activate STZ transitions.  But we do not
need to consider such situations here, and therefore $\zeta \to \infty$
will suffice.  In this limit, ${\cal C}(\tilde s)
\approx \Theta(|\tilde s|-1)\,(|\tilde s|-1)$, where $\Theta$ is the
Heaviside step function, and
\begin{equation}
\label{q0}
q_0(\tilde s) \approx \Theta(|\tilde s|-1)\,\frac{(|\tilde s|-1)(\tilde s ~{\rm sign}(\tilde s)-1)}{\tilde s}
\end{equation}

The initial conditions in Eqs.(\ref{initial})
must be supplemented with an initial condition for the effective temperature:
\begin{equation}
\chi(r,t=0) = \chi_0 \ , \label{initial_1}
\end{equation}
where $\chi_0$ describes a homogeneous state of disorder of the material
before any load is applied. Because we are using Eq.(\ref{m0}), there
is no comparable initial condition for $m$.

Eqs.(\ref{Force_Bal_2}), (\ref{kinematic1})-(\ref{kinematic2}),
(\ref{BC_v3})-(\ref{v_W}), (\ref{STZ1})-(\ref{q0}), and the initial
conditions in Eqs.(\ref{initial}) and (\ref{initial_1})  define our problem.
This system is specified by the following set of dimensionless
parameters: $\epsilon_0, \quad c_0, \quad \chi_{\infty},\quad \chi_0, \quad
\mu/s_y, \quad K/s_y$ and a given loading scheme
$\sigma^{\infty}(t)$. The first three parameters characterize our plasticity theory;
$\chi_0$ specifies the initial state of disorder of the material; and the last two
parameters are the elastic moduli in units of the yield stress $s_y$.
Note that two parameters, $\tau_0$ and $s_y$, are scaled out as units of time and
stress respectively.  The only length scale in the problem is the initial
radius of the hole which, without loss of generality, we set to unity.

\section{Loading Schemes and the Incompressible limit}
\label{loading}
\subsection{Loading schemes}

We now specify the loading scheme
$\sigma^{\infty}(t)$. To make contact with earlier work \cite{99LL}, we
will look briefly in what follows at the case of increasing the load to a constant value, that is,
\begin{equation}
\label{const-load}
\sigma^{\infty}(t) = \cases{\sigma_0 \frac{t}{500 \tau_0},&for $0<t<500\tau_0$\cr \sigma_0,&for $t>500\tau_0$.}
\end{equation}
Our primary interest here, as in \cite{99LL}, will be to see how the region
of plastic deformation forms, expands, and reaches a stable, stationary state
for loading stresses $\sigma_0$ less than the threshold for unbounded growth
of the hole (see Appendix \ref{cavitation}).

We will pay greater attention to a class of situations in which the load
takes the form of a stress pulse of finite duration, because we believe that such
a pulse may be relevant to fracture dynamics.  Plastic deformation
of a material failing by crack propagation is localized (if it happens
at all) near the fracture surfaces. At any given material point, deformation
occurs only during a short time interval as the crack tip passes nearby.
Specifically, a material element lying ahead of a crack tip experiences
first an increasing stress as the crack tip opens, and then a decreasing
stress as the load vanishes on the newly formed -- and deformed --
fracture surfaces.

\begin{figure}
\centering
\epsfig{width=.45\textwidth,file=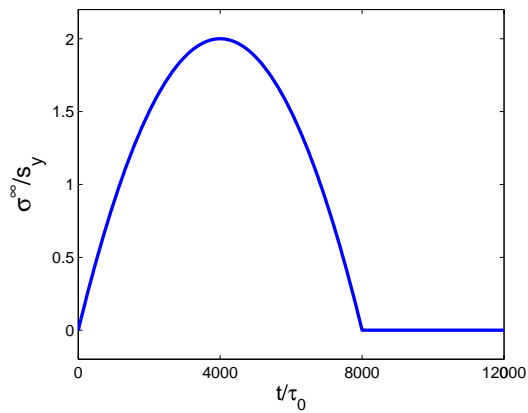}
\caption{A time dependent non-monotonic loading scheme corresponding
to Eq. (\ref{non-monot}) with $T=8000\tau_0$ and $\sigma_p=2s_y$.}
\label{load_parabolic}
\end{figure}

To simulate such a process in the circular geometry, we study a time
dependent loading scheme in which the remote stress $\sigma^{\infty}(t)$
increases monotonically from zero to a peak stress $\sigma_p > s_y$, and
then returns monotonically to zero:
\begin{equation}
\sigma^{\infty}(t)=\left\{\begin{array}{r@{, \quad}l}
4\,\sigma_p\,\frac{\displaystyle t(T-t)}{\displaystyle T^2} & 0<t<T,
\\ 0 & t>T. \end{array} \right.
\label{non-monot}
\end{equation}
An example with $T=8000\,\tau_0$ and $\sigma_p=2\,s_y$ is shown in Fig.
\ref{load_parabolic}.

Our choice of $T$ in Eq.(\ref{non-monot}) is an order-of-magnitude
estimate emerging from the analysis in \cite{00Lan}.  For cracks
whose speeds are governed by plastic dissipation, and
when surface tension is negligible, that analysis implied that
the crack tip blunts in such a way that the stress on its surface
is always of order $s_y$, and the tip radius is
linearly proportional to the crack speed.  Thus the time $T$ that
any material element spends in the fracture zone is roughly constant,
of order the plastic relaxation time $\tau_0\exp{(1/\chi_{\infty})}$.
If $\chi_{\infty} \cong 0.13$, then  $T\cong 2200\,\tau_0$.  This estimate
is consistent with the discussion in \cite{06BPP}, where the plastic
zone size was dynamically adjusted so that the time needed for the
crack to pass this zone was of the order of the plastic relaxation time.
Note that $T$ is a very short time, of order picoseconds if
$\tau_0$ is an atomic time of order femtoseconds.  That value of $T$
is roughly an atomic spacing divided by a sound speed. These estimates
lead us to believe that, although STZ plasticity is slow on atomic time scales,
it is easily fast enough to be relevant to fracture.  Clearly, this estimate
of relevant time scales will have to be reexamined when the theory is applied
to realistic models of fracture where surface tension is important.
\subsection{The incompressible limit}

It is convenient for both analytic and numerical simplicity to work in
the limit of elastic (as well as plastic) incompressibility.
Many materials are highly incompressible, as
evidenced by large bulk moduli $K$ or, equivalently, by
the two-dimensional Poisson ratios $\nu^*$ being close to unity.
Moreover, we have studied numerically the set of equations
(\ref{Force_Bal_2}), (\ref{kinematic1})-(\ref{kinematic2}) and
(\ref{W})-(\ref{pdf}) for various loading scenarios using
$0.7\le\nu^*\le1$, and have found results that vary smoothly
with $\nu^*$ without any noticeable singular behavior in the limit
$\nu^*\to 1$. Therefore, for analytic purposes, we can focus on the
limit $\nu^*\!\to\! 1$ ($K\! \to\! \infty$).

Note first that, in the incompressible limit, Eq. (\ref{kinematic1}) reduces to
\begin{equation}
\frac{v}{r}+\frac{\partial v}{\partial r} = 0 \label{incomp} \ ,
\end{equation}
which is in accord with the zero total compressibility $tr\B D^{tot}=0$. Eq.
(\ref{incomp}) can immediately be integrated to yield
\begin{equation}
v(r,t) = \frac{\dot{R}(t)R(t)}{r} \ , \label{incomp_v}
\end{equation}
where the boundary condition Eq. (\ref{edge_velocity}) has been used.
This explicit expression for the velocity field provides a major
simplification because we no longer need to solve
Eq. (\ref{v_W}) numerically.

Having the velocity field at hand, we
can rewrite all the equations of the theory in a more explicit way.
The sum of Eqs. (\ref{kinematic1}) and (\ref{kinematic2}) becomes:
\begin{equation}
\frac{\dot{R}R}{r^2}=\frac{1}{2\mu}\frac{\partial s}{\partial t}+
\frac{1}{2\mu}\frac{\dot{R}R}{r}\frac{\partial s}{\partial r}+D^{pl}
\ . \label{first_eq}
\end{equation}
Eqs.(\ref{STZ1})-(\ref{q0}) (with $\tilde s =s/s_y$ as before) are:
\begin{eqnarray}
&&D^{pl} = {\epsilon_0\over\tau_0}\,e^{-1/\chi}\,q_0(\tilde s),~~~~ q_0(\tilde s) =  \Theta(|\tilde s|-1)\,\frac{(|\tilde s|-1)(\tilde s ~{\rm sign}(\tilde s)-1)}{\tilde s}
;\label{STZ1_incom}\\
&&\frac{\partial \chi}{\partial t} + \frac{\dot{
R}R}{r}\frac{\partial \chi}{\partial r} = {2\,\epsilon_0\over
c_0\,\tau_0}\,e^{-1/\chi}\,\tilde s\,q_0(\tilde s)\,(\chi_{\infty}-\chi) \ . \label{STZ3_incom}
\end{eqnarray}
The derived relation, Eq.(\ref{BC_v3}), becomes:
\begin{equation}
\frac{\dot{R}}{R} = \frac{\displaystyle 2\int_{R}^{\infty}
\frac{D^{pl}}{r}dr +\frac{\dot{\sigma}^{\infty}(t)}{2\mu}}
{\displaystyle 1-\frac{s(R,t)}{\mu}} \label{R_dot_incom}.
\end{equation}
Note that the pressure $p(r,t)$ does not appear in
the final equations for the incompressible limit, but is
computable as before from the deviatoric stress field $s(r,t)$ using Eq.
(\ref{Force_Bal_2}).


\section{Results}
\label{results}

We are now ready to study numerically Eqs.
(\ref{first_eq})-(\ref{R_dot_incom}) in detail. We consider both the
constant load and the time dependent loading scenarios,
Eqs.(\ref{const-load}) and (\ref{non-monot}) respectively.
\begin{figure}
\centering \epsfig{width=.9\textwidth,file=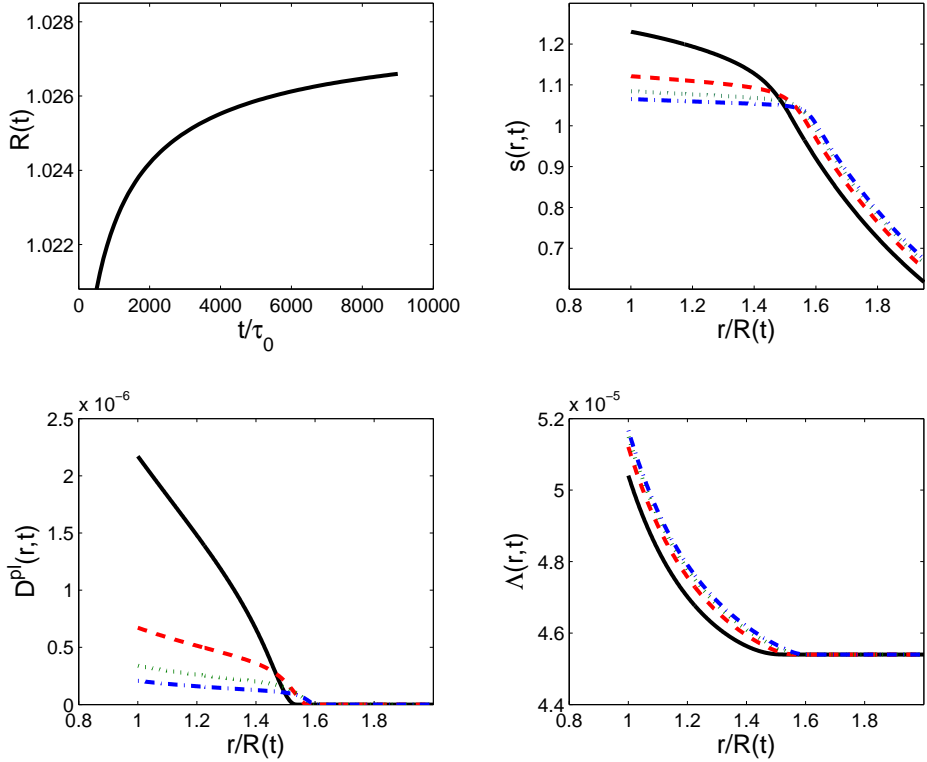}
\caption{The $r$ dependence of the dynamical variables at different times in the constant load case. The radius of the circle is shown in the upper left panel without the initial elastic expansion.  At longer times the radius saturates. Note the relative small increase in radius. The stress is shown in the upper right panel, and the fields $D^{pl}(r,t)$ and $\Lambda=e^{-1/\chi}$ in the two lower panels respectively.  The parameters chosen are $\epsilon_0=1, \quad c_0=1, \quad \chi_{\infty}=0.13, \quad \chi_0=0.1, \quad \mu=50\,s_y$ and $\sigma^{\infty}=2s_y$. The solid line corresponds to $t=2000\tau_0$, the dashed line to $t=4000\tau_0$, the dotted line to $t=6000\tau_0$ and the dotted-dashed line
to $t=8000\tau_0$.}  \label{cons_stress}
\end{figure}

\subsection{Increase toward a constant load}

The results of numerical simulations for the case of an increase toward a constant load,
$\sigma_0/s_y = 2$, are shown in Fig. \ref{cons_stress}. In this
loading scheme, we know that there is a maximum
value of $\sigma^{\infty}$, say $\sigma^{th}$, above which the
hole expands indefinitely. (See \cite{99LL} and earlier references cited there.)
For completeness, in Appendix \ref{cavitation}, we show how to compute
$\sigma^{th}$ in the current version of STZ theory.  The result, for
the parameters used here, is $\sigma^{th} \approx 5 s_y$, which is substantially
larger than the value of $\sigma_0$ chosen for this illustration.

Note that the radius of the hole, $R(t)$, first increases elastically to
$R_0 = \exp (\sigma_0/2\,\mu)\cong 1.02$ (not shown in Fig. \ref{cons_stress}),  then grows (by less than one percent in this example), and tends toward a constant value
at large times.  The stress $s$ is proportional to the elastic solution,
$1/r^2$, for $r$ outside the plastically deformed region; but,
in accord with conventional plasticity theories, $s$
decreases in time toward the yield stress $s_y$ inside that region.
The plastic region eventually extends
out to a radius $R_1$.  The usual quasistatic estimate of $R_1$
\cite{Lubliner,99LL} is made by assuming $s(r) = s_y$ for $R<r<R_1$,
and $s(r) = s_y\,(R_1/r)^2$
for $r > R_1$.  Inserting this $s(r)$ into the integral on the
right-hand side of Eq.(\ref{integral}) and setting
$\sigma^{\infty} = \sigma_0 = 2\,s_y$ on the left-hand side, we find
$R_1/R = \exp[(1/2)\,(\sigma_0/s_y -1)] = 1.65$. Thus,  even in cases
where the plastic deformation is small in comparison to the
elastic displacement, the plastically deformed region may be
quite extensive. Finally, note that the
STZ density $\Lambda=e^{-1/\chi}$ becomes substantially larger
than its initial value throughout this plastic region.
\subsection{Time dependent non-monotonic loading}

\begin{figure}
\centering \epsfig{width=.9\textwidth,file=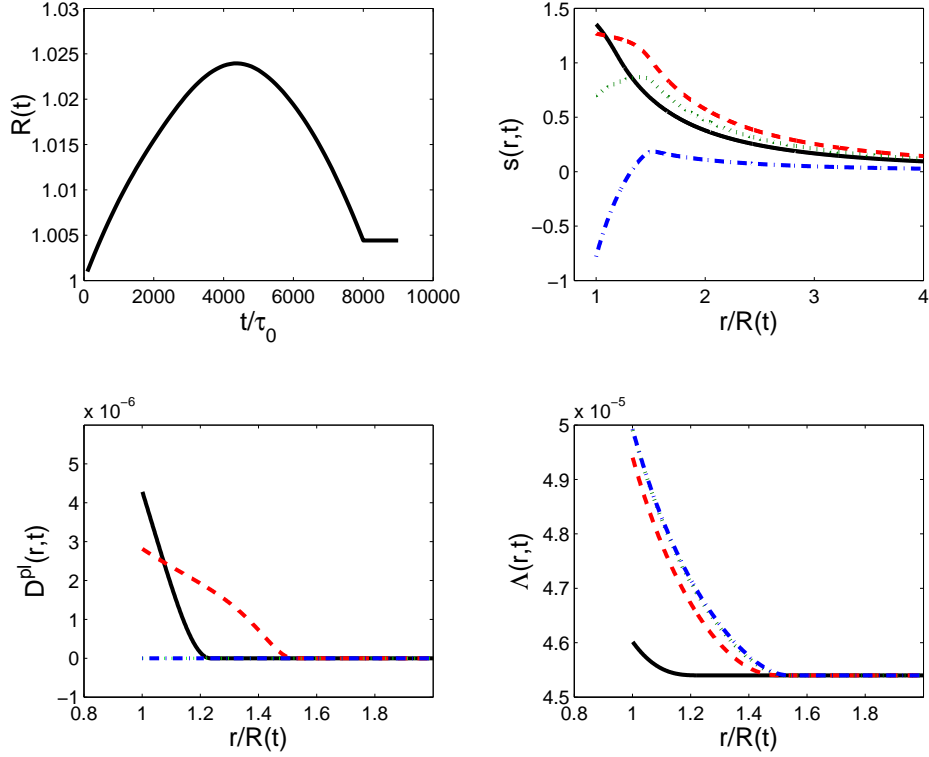} \caption{The time dependent non-monotonic load case. All the material parameters are the same as those for Fig. \ref{cons_stress}, while the load is given by  Eq. (\ref{non-monot}) with $T=8000\tau_0$ and $\sigma_p=2 s_y$. The solid line corresponds to $t=2000\tau_0$, the dashed line to $t=4000\tau_0$ the dotted line to $t=6000\tau_0$ and the dotted-dashed line to $t=8000\tau_0$.} \label{load_unload2} \end{figure}

\begin{figure}
\epsfig{width=1.0\textwidth,file=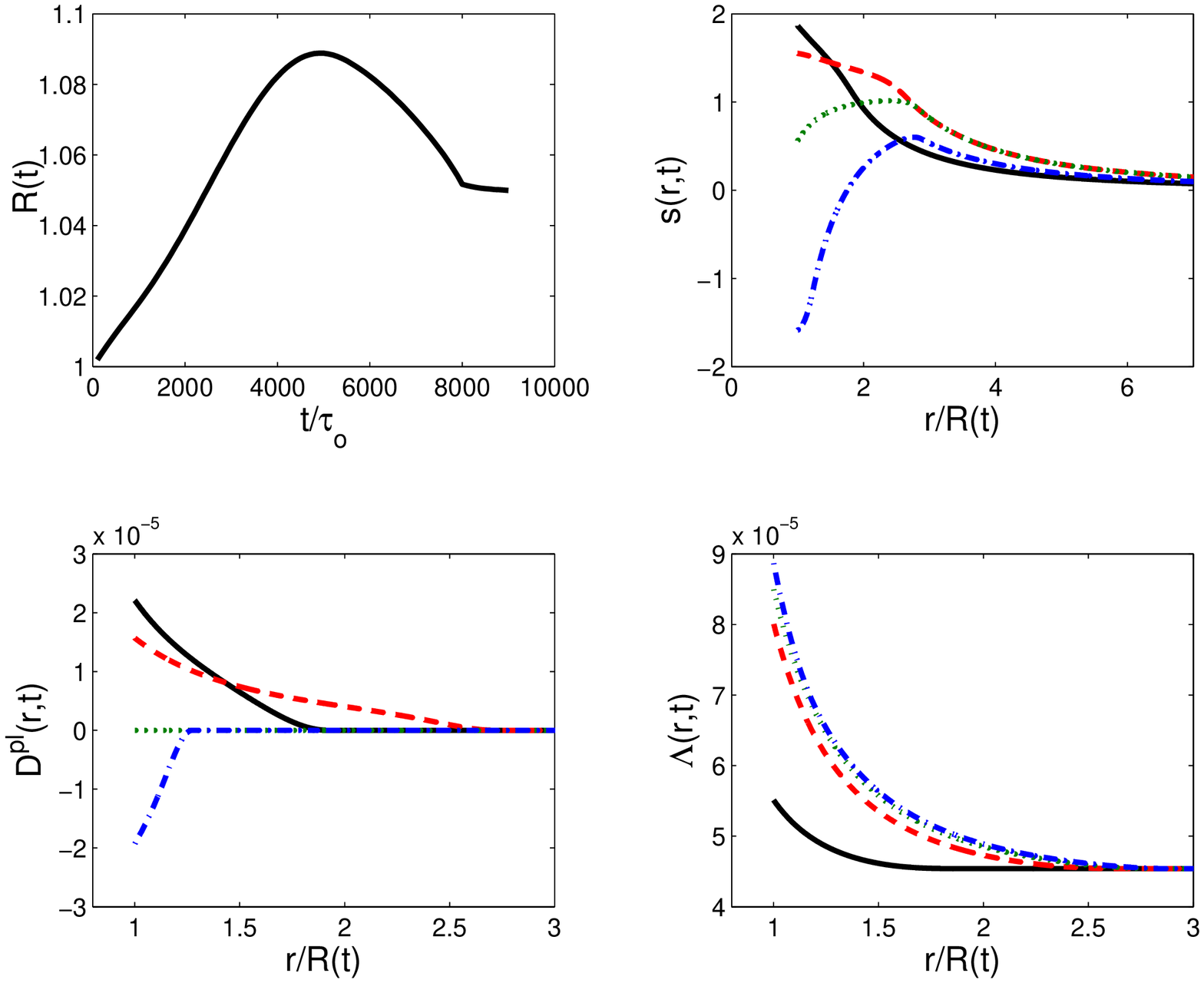} \caption{Same as preceding figure but with $\sigma_p=4 s_y$.} \label{load_unload4}
\end{figure}

In Figs. \ref{load_unload2} and \ref{load_unload4} we present numerical results for the case of
time-dependent non-monotonic loading as shown in Eq. (\ref{non-monot})
and Fig. \ref{load_parabolic}. Here we have chosen $T=8000\,\tau_0$ and
two different values of the peak stress, $\sigma_p/s_y = 2.0$ and $4.0$.
The first of these is small enough that the radius of the hole changes
at its maximum by only about two percent, consistent with the effect of
constant loading at the same stress shown in Fig. \ref{cons_stress}.
Only a small part of that transient change in the radius is irreversible
plastic deformation; most of what we are seeing is elastic expansion and
contraction of the system as a whole.  However, this relatively moderate scenario
may be the more realistic of the two sets of time dependent simulations for
describing failure in strong structural materials where, in keeping with
conventional analyses, we do not expect stresses to be much larger than the
yield stress.

The bigger peak stress causes substantially more plastic deformation,
as seen in the comparatively large, irreversible change in the hole radius
that remains after the system has been fully unloaded.  This
peak stress may be unrealistically large; but showing it here
illustrates some features of the plastic response more clearly
than the small-stress case.  Note that the
large applied stress induces a complex unloading sequence in which
the stress near the hole becomes so large and negative that it
drives strain recovery {\it via} reverse plastic deformation.

The dynamic response of this system, in both examples, is characterized
by the emergence of a residual stress in the plastic zone near the hole
edge. Note that $s$ exhibits the sharp transition between plastic and
elastic regions that we expect as a result of the special form of the
constitutive law chosen in Eq.(\ref{q0}), and that the STZ density
$\Lambda$ becomes large in the plastic region.  The qualitative picture
that we take from these results is that of a  plastically deformed region
near the hole edge
(or near any defect boundary in the general case), outside of
which $s$ retains its elastic solution proportional to $1/r^2$,
but within which $s$ decreases and may change sign, consistent with
an irreversible outward displacement of the material near the hole.

\section{ Boundary-Layer Approximation}
\label{BLsect}

A principal objective of this investigation has been to study
the dynamics of time-dependent plasticity in the neighborhood of a
symmetrically loaded circular hole, and, from this study, to learn how
to construct realistic approximations for less symmetric situations.
We have seen in the preceding Section that, under certain circumstances,
plastic deformation remains localized in a relatively narrow zone near the hole.
This observation leads us to believe that some sort of boundary-layer
approximation might have at least limited validity in interesting
applications such as asymmetric void growth or fracture.  In what follows,
we explore one class of such approximations.

The numerical examples illustrated in Figs. \ref{load_unload2} and \ref{load_unload4} have several common
features that need to be captured in any useful approximation.
Most importantly, the dynamic behavior is qualitatively
different in what we call the ``active plastic phase'' -- roughly speaking,
the time interval in which the hole is growing by plastic deformation
 -- than it is in the ``elastic unloading phase'' where plastic deformation ceases.
During the active plastic phase, as soon as $\sigma^{\infty}$ exceeds $s_y$, plastic deformation
occurs in an active plastic zone, $R(t) < r < R_1(t)$, where $R_1(t)$ is
the radius at which
\begin{equation}
\label{R1}
s\bigl(R_1(t),t\bigr) = s_y.
\end{equation}
During this phase, no irreversible deformation has occurred for $r > R_1(t)$,
and thus the stress field in this outer region is simply
\begin{equation}
s(r,t)=s_y \frac{R_1^2(t)}{r^2}, \quad p(r,t)=-\sigma^{\infty}(t),
\quad \hbox{for} \quad r>R_1. \label{elastic_R1}
\end{equation}

This behavior changes as elastic unloading begins. As seen
in Figs. \ref{load_unload2} and \ref{load_unload4},  $R(t)$ reaches its peak considerably later than
when the load $\sigma^{\infty}(t)$ peaks at $t_p= 4000\,\tau_0 $.  Between $t_p$ and the time at
which $R(t)$ reaches its maximum, and for some time after that, the hole
continues to grow plastically while the system as a whole contracts
elastically.  During that interval, the stress at the boundary of the hole,
$s(R,t)$, drops rapidly toward the yield stress $s_y$; while $s(R_1,t)$, by
the definition of $R_1$, remains equal to $s_y$. The transition between
the active plastic  and the elastic unloading phases that is important for our purposes
occurs when $s(R,t)$ passes downward through $s_y$ at a time that we call $t_1$.

\begin{figure}
\centering \epsfig{width=.9\textwidth,file=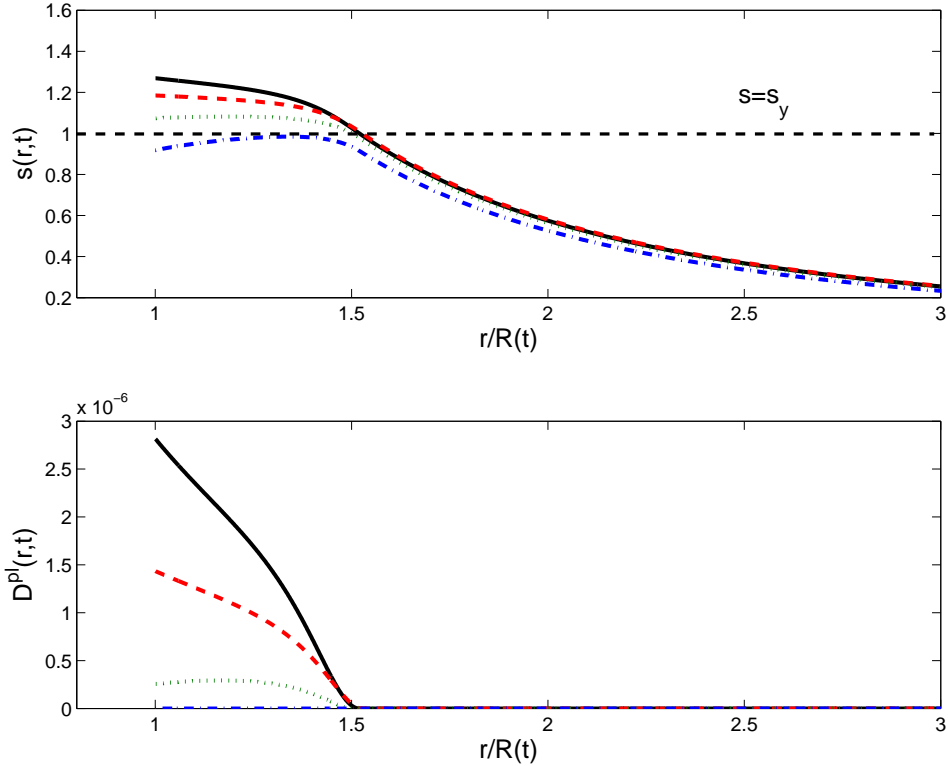}
\caption{The
profiles of $s(r,t)$ and $D^{pl}(r,t)$ after the external stress
the same as those for Fig. \ref{load_unload2}. The solid line
corresponds to $t=4000\tau_0$, the dashed line to $t=4500\tau_0$
the dotted line to $t=5000\tau_0$ and the dotted-dashed line to
$t=5500\tau_0$.} \label{zoom}
\end{figure}

To see what happens near $t_1$, we show in Fig. \ref{zoom}, for the case
$\sigma_p = 2s_y$, a sequence of graphs of $s(r,t)$ and $D^{pl}(r,t)$
as functions of $r/R(t)$  for four equally spaced times starting
at $t_p = 4000\,\tau_0$, and ending at $t= 5500\,\tau_0$ (just later
than $t_1\cong 5200\,\tau_0$).  Note that at $t=5500\,\tau_0$ $D^{pl}$ has
already vanished.  The crucial observation is that
$D^{pl}(r,t)$ extends across the whole active plastic zone, $R(t)<r<R_1(t)$,
until $t=t_1$, at which time, at least to a rough first
approximation, it vanishes almost uniformly everywhere.

The elastic unloading phase, starting at about time $t_1$, evolves as
an entirely elastic response to the decreasing driving force, conditioned
by the material displacements that occurred during the plastic loading phase.
During this part of the unloading process, there is no longer
an active plastic zone, but there remains a plastically deformed region --
a ``process zone'' -- within which irreversible material displacements give
rise to residual stresses.  The process zone extends from $R(t)$ out to the
most distant material point at which plastic deformation occurred, that is,
out to $R_1(t_1)$ advected back toward $R(t)$ by elastic relaxation.

Elastic unloading persists at least until $t = T$ when the remote
load $\sigma^{\infty}(t)$ vanishes.  As seen in Fig. \ref{load_unload4}, the stress near
the hole may become so large and negative that it drives plastic strain
recovery.  This behavior occurs only for large peak stresses, and does
not occur when the remote stresses are smaller than about $2.5\,s_y$
in the present model.  We will not include plastic strain
recovery in our boundary-layer analysis, and simply will point out where it
is missing.

Our strategy for developing a boundary-layer approximation is to write
equations of motion for quantities defined only near the boundary,
i.e. within the process zone, and then to deduce
the boundary motion from these local quantities instead of solving for
fields everywhere in the system.  In our case, the relevant boundary
is just the hole radius $R(t)$.
The rate of change of any quantity $A(R,t)$, defined
on $r\!=\!R(t)$, is
\begin{equation}
\frac{dA(R,t)}{dt} = \frac{\partial A(R,t)}{\partial t} + \dot{R}
\frac{\partial A(R,t)}{\partial r} \ . \label{der_bound}
\end{equation}
Since $\dot{R}=v(R,t)$, the time derivative in Eq. (\ref{der_bound})
is just the material time derivative defined in Eq. (\ref{Mat_Der}).
The equations for $s(R,t)$  and $\Lambda(R,t)$, {\it i.e.} Eqs. (\ref{first_eq} - \ref{STZ3_incom}) evaluated at $r = R(t)$, are:
\begin{eqnarray}
&&\frac{\dot{R}}{R}= \frac{1}{2\mu}\frac{d s(R,t)}{d t}+ D^{pl}(R,t); \label{final eqs_bl1}\\
&&D^{pl}(R,t) = {\epsilon_0\over\tau_0}\,e^{-1/\chi(R,t)}\,q_0(R,t);~~~q_0(R,t)\equiv q_0\bigl(\tilde s(R,t)\bigr); \label{final eqs_bl2}\\
&&\frac{d \chi(R,t)}{d t} =
{2\,\epsilon_0 \over
c_0\,\tau_0}\,e^{-1/\chi(R,t)}\,s(R,t)\,q_0(R,t)\,\left[\chi_{\infty}-\chi(R,t)\right]\
. \label{final eqs_bl4}
\end{eqnarray}

We need one more relation to close this set of equations.  The
obvious candidate is Eq.(\ref{R_dot_incom}), which
already involves only quantities defined in the process zone.
This equation demands special attention, however, because it
has been derived using exact mathematical
relationships that are unique to circular symmetry,
and because it explicitly involves the remote driving force
$\sigma^{\infty}(t)$ which entered the analysis {\it via}
those exact relationships.  We need some such relation to determine
the coupling between the remote driving force and the boundary.
If the boundary-layer strategy is to be successful, we ultimately
will have to interpret $\sigma^{\infty}(t)$ as a
quantity emerging from solutions of a more general elasticity
problem. For present purposes, however, we accept
Eq.(\ref{R_dot_incom}) as written, except that we simplify it
by neglecting the term $s(R,t)/\mu \approx s_y/\mu \ll 1$ in
the denominator on the right-hand side.  Thus we write:
\begin{equation}
\frac{\dot{R}}{R} \cong 2\int_{R}^{R_1}
\frac{D^{pl}(r,t)}{r}\,dr+\frac{\dot{\sigma}^{\infty}(t)}{2\mu}
\ . \label{third_eq_simple_2}
\end{equation}
Note that the original integral in Eq. (\ref{R_dot_incom})
involves the plastic rate of deformation $D^{pl}(r,t)$ which
vanishes for $r\!>\!R_1$. Therefore, in Eq. (\ref{third_eq_simple_2}),
we have explicitly inserted $R_1$ as the upper limit of integration.

Next we use
Eq.(\ref{third_eq_simple_2}) to obtain an approximate expression
for $\dot R/R$, and then use that relation to eliminate
$\dot R/R$ on the left-hand side of Eq.(\ref{final eqs_bl1}) in favor of
$s(R,t)$.  To do this, we need
an approximation for the integral over $D^{pl}(r,t)$ that appears in
Eq.(\ref{third_eq_simple_2}).  The numerical results shown in
Fig. \ref{zoom} suggest that,
throughout the active plastic phase, $D^{pl}(r,t)$ varies
almost linearly, from its value $D^{pl}(R,t)$
at $r=R(t)$ to zero (by definition) at $r = R_1(t)$.  Thus we write
\begin{equation}
\label{Dintegral}
\int_R^{R_1} {D^{pl}(r,t)\over r}\,dr \cong {1\over 2}\,D^{pl}(R,t)\,\left({R_1(t)\over R(t)}-1\right).
\end{equation}
To estimate $R_1(t)$ we invoke
Eq.(\ref{integral}), which is another exact relationship between
the circularly symmetric stress field $s(r,t)$ and the remote
driving force $\sigma^{\infty}(t)$, and is subject to the same
concerns that we expressed above about our use of Eq.(\ref{R_dot_incom}).
So long as we remain in the active plastic phase, and can assume that $R_1(t)$ remains
the outer boundary of the material region in which $D^{pl} (r,t)$ is nonzero,  we can use Eq.(\ref{elastic_R1})
for the stress outside $r = R_1(t)$, and write Eq.(\ref{integral})
in the form
\begin{equation}
\label{integral2}
\sigma^{\infty}(t)= 2\,\int_{R(t)}^{R_1(t)}{s(r,t)\over r}\,dr + s_y.
\end{equation}
Then, in the spirit of our approximation for $D^{pl}(r,t)$ in the active
plastic region, we make a linear approximation for $s(r,t)$,
using the fact that this stress is equal to $s_y$ at $R_1$, and allowing
$s(R,t)\equiv s\bigl(R(t),t\bigr)$ to be an as yet undetermined function
of time.  Specifically, for $R(t)<r<R_1(t)$ and for times $t<t_1$ such that
$s(R,t) \ge s_y$, we have:
\begin{equation}
\label{linear-s}
s(r,t)\cong s_y + \frac{R_1(t)-r }{ R_1(t)-R(t)}(s(R,t)-s_y)\ , \quad s(R,t)> s_y.
\end{equation}
Inserting Eq.(\ref{linear-s}) into Eq.(\ref{integral2}), and
linearizing in $(R_1-R)/R$, we find, again for $t<t_1$,
\begin{equation}
\label{thin boundary}
\left({R_1(t)\over R(t)}-1\right)\cong \frac{\sigma^{\infty}(t)- s_y}{s(R,t)+s_y}\ ,  \quad \sigma^{\infty}(t),\,s(R,t)> s_y.
\end{equation}
Note that when $\sigma^{\infty}(t)$ is increasing and is smaller than $s_y$, there is no
active plastic zone at all, i.e. $R_1(t)=R(t)$.
Using Eq.(\ref{Dintegral}), Eq.(\ref{third_eq_simple_2}) becomes
\begin{equation}
\frac{\dot{R}}{R} \cong  \frac{\sigma^{\infty}(t)- s_y}{s(R,t)+s_y} D^{pl}(R,t)
+\frac{\dot{\sigma}^{\infty}(t)}{2\mu} \ ,  \quad \sigma^{\infty}(t),\,s(R,t)> s_y
\ . \label{fourth_eq_simple_2}
\end{equation}

We emphasize that, in this approximation, the end of the active plastic
phase occurs at the time $t_1$ when $s(R,t_1)=s_y$.  By definition,
$D^{pl}(R,t_1)=0$, and therefore $D^{pl}(r,t_1)\cong 0$ throughout
$R(t_1)<r<R_1(t_1)$.
Apart from the possibility of reverse plasticity, $D^{pl}(r,t)$ vanishes
at all later times, $t>t_1$.  It follows that Eq. (\ref{fourth_eq_simple_2}),
with a properly interpreted  $D^{pl}(r,t)$, is a valid approximation
at all times, during loading and unloading, because
-- apart from the possibility of reverse plasticity --
$D^{pl}(R,t)$ vanishes whenever $s(R,t) < s_y$.

\begin{figure}
\centering \epsfig{width=1.0\textwidth,file=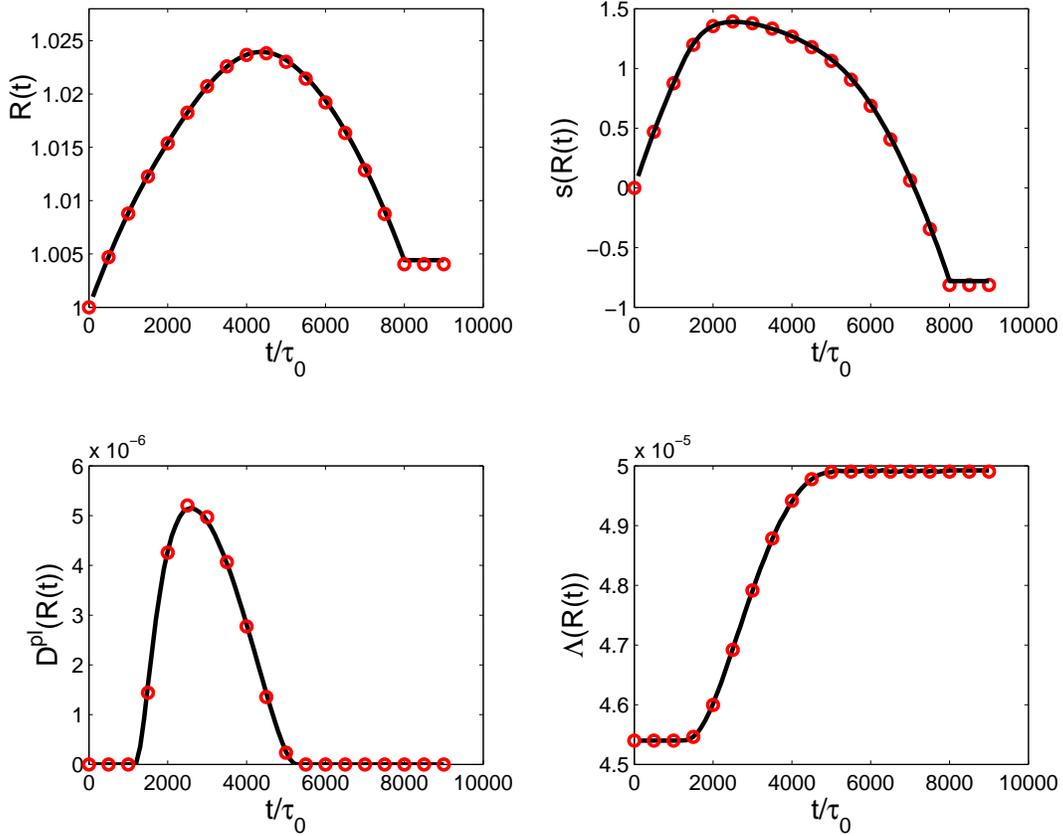}
\caption{A comparison between the exact solution (solid line) and the prediction of the boundary layer
theory (open circles) for $\sigma_p=2s_y$.}
\label{BL1}
\end{figure}

\begin{figure}
\centering \epsfig{width=1.0\textwidth,file=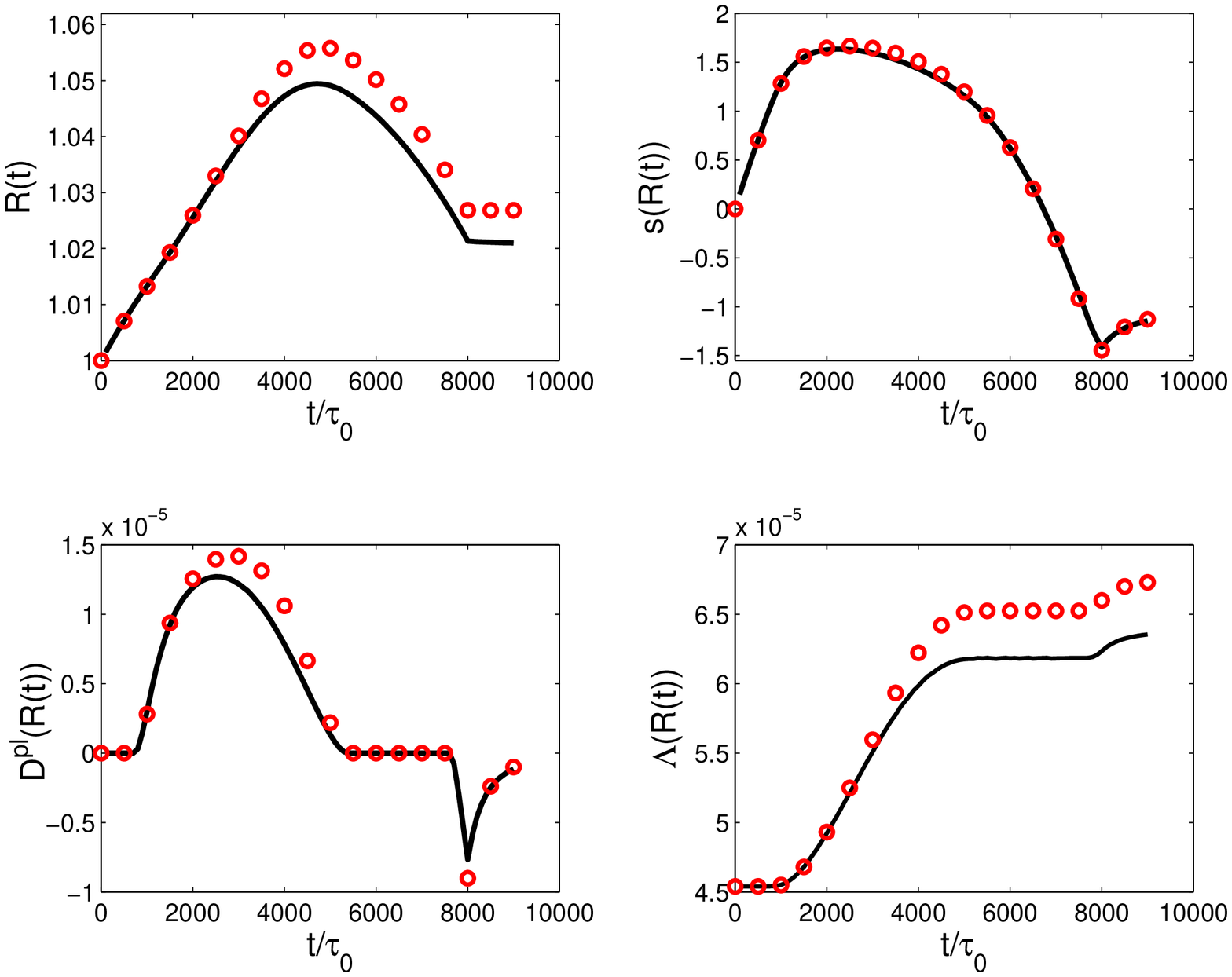}
\caption{A comparison between the exact solution (solid line) and the prediction of the boundary layer
theory (open circles) for $\sigma_p=3s_y$.}
\label{BL2}
\end{figure}

\begin{figure}
\centering \epsfig{width=1.0\textwidth,file=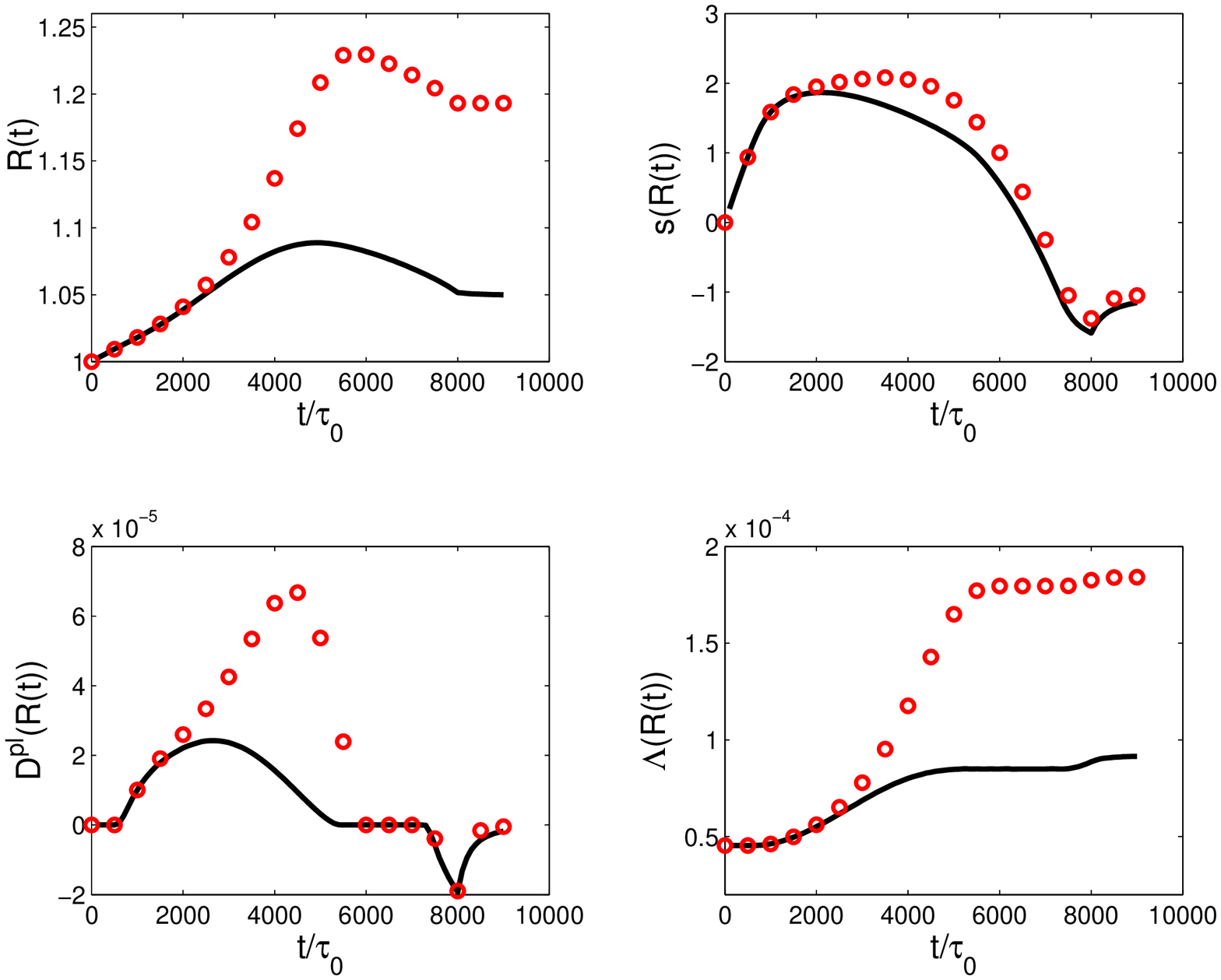}
\caption{A comparison between the exact solution (solid line) and the prediction of the boundary layer
theory (open circles) for $\sigma_p=4s_y$.}
\label{BL3}
\end{figure}

Our boundary-layer calculation is straightforward from here on.
We insert Eq. (\ref{fourth_eq_simple_2}) on the left-hand side of
Eq. (\ref{final eqs_bl1}), thus obtaining a nonlinear, first-order
differential equation for $s(R,t)$, which can be solved using
Eqs. (\ref{final eqs_bl2})-(\ref{final eqs_bl4}).  Our results are
shown in Figs. \ref{BL1}-\ref{BL3}, where we compare the predictions of the boundary layer theory
for $R(t)$, $s(R,t)$, $D^{pl}(R,t)$ and $\Lambda(R,t)$ with the exact solutions for
$\sigma_p/s_y = 2$, 3 and 4.
The comparison is excellent for $\sigma_p/s_y = 2$, even though $R_1(t)/R(t) -1$ is about 0.5, which is
not that small. The results are acceptable also for  $\sigma_p/s_y = 3$, but deviations grow rapidly for higher values of
 $\sigma_p/s_y$, as expected. Note that in Figs. \ref{BL2} and \ref{BL3} the effect of plastic
 strain recovery is noticeable for the exact solution but has not been included in the boundary-layer approximation.
 Nevertheless, the values $s(R(t))<-s_y$ and $D^{pl}(R(t))<0$ are well approximated in this period of
 plastic strain recovery.

\begin{figure}
\centering \epsfig{width=1.0\textwidth,file=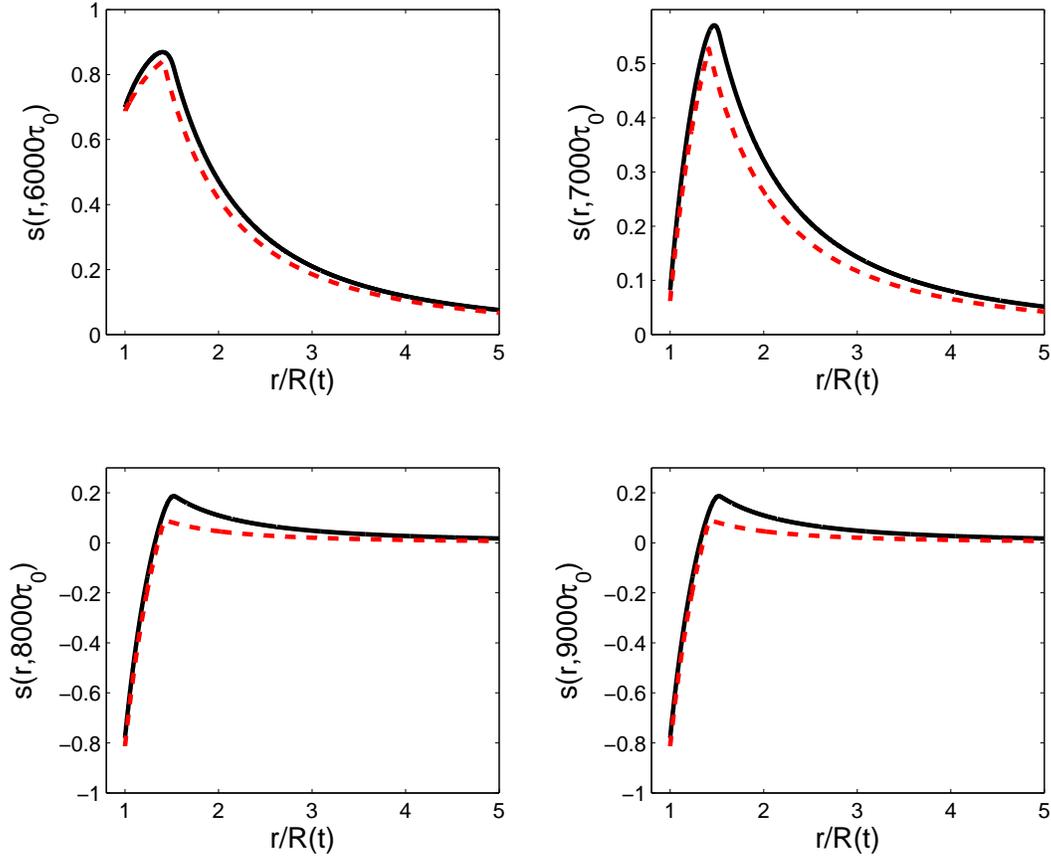}
\caption{Comparison between the full solution (solid line) for $s(r,t)$ and the prediction of the boundary layer
approximation (dashed line) for $\sigma_p = 2\,s_y$ and for various values of $t$.}
\label{Jim}
\end{figure}

We can push the comparison between the boundary-layer approximation and the
exact results further by computing the residual stresses in the process zone.
We do this as follows.  We note first that Eq.(\ref{final eqs_bl1})
can be used for any radius $R'(t)$, not just the radius of the hole $R(t)$.
At times later than $t_1$, the quantity $D^{pl}(R',t)$ appearing here vanishes for any $R'$.
We also can compute the quantity $\dot R'/R'$ on the left-hand
side of Eq.(\ref{final eqs_bl1}) using our knowledge of $\dot R(t)/R(t)$ and the
fact that the material in the region between $R$ and $R'$ is incompressible
($R'\,\dot R'=R\,\dot R$).  Finally, we know that $s(R',t_1)\cong s_y$ for all
$R'$ between $R$ and $R_1$; so we can
use Eq.(\ref{final eqs_bl1}) to compute the advected value of $s\big(R'(t),t\Big)$
at any later time $t$, and then plot this value as a function of the advected
position $r=R'(t)$.  The functions $s(r,t)$ computed in this way, for
$\sigma_p/s_y = 2$, are shown in Fig. \ref{Jim}. Outside the process zone, we have used the elastic solution with $s \propto 1/r^2$.
The agreement between the full solution and the boundary layer predictions lends further support
to the philosophy presented in this section.
\section{Discussion and Summary}
\label{summary}

We have presented a derivation of the equations of motion for
a moving free boundary, in this case a circle, using linear elasticity
and the athermal STZ theory of amorphous plasticity. These equations
were solved numerically and yielded some interesting results. We looked
especially at loading scenarios in which the material was subjected, at
large distances from the hole, to stress pulses whose durations were
comparable to the plastic relaxation time for the STZ mechanism.
For the range of material parameters chosen here, which we believe to be characteristic
of realistic amorphous solids, the irreversible displacement of the boundary
of the hole was small.  However, the width of the region in which plastic
deformation occurred, as evidenced by increased internal disorder and residual
stresses, became comparable to the radius of the circle for stress pulses whose
peak strengths $\sigma_p$ were twice the yield stress $s_y$. We also found that strong
enough stress pulses, $\sigma_p \ge 2.5\,s_y$ induce reverse, plastic strain
recovery near the edge of the hole.

The theory as written is rather cumbersome even for the
circular symmetry considered here. It is therefore with some relief that we
have found that a boundary layer approximation, using only quantities defined
on the circumference of the circle and their coupling to the remote stress,
succeeds in capturing the plastic effects quite well. Specifically, despite
several apparently serious oversimplifications, this approximation accurately
reproduces the irreversible displacement of the radius of
the circle and the residual stress found in its neighborhood.
We propose that such boundary layer theories might find
useful applications in less symmetric situations, not the least interesting
being fracture dynamics.

A next issue on our agenda is the dynamic stability of our circular solutions.
It will be interesting to learn whether our equations of motion predict that
the growing circular hole becomes unstable against symmetry-breaking perturbations,
for example, whether it forms fingers that might evolve into cracks,
or whether a perturbed circular solution can survive. The transition
between these two possibilities, if found, might shed light on the
brittle-to-ductile transition in visco-plastic materials. Along the same lines,
it will be important to learn how strongly the circular behavior itself,
and the stability of circular symmetry, depend on specific details of our
model such as elastic compressibility, initial states of disorder, or the absence
of thermal effects.

\acknowledgments

We thank A. Lemaitre for illuminating discussions. This work was
supported in part by the Minerva Foundation with funding from the Federal German Ministry for Education and Research, the Israel Science Foundation and the German Israeli Foundation.
E.B. was supported by a Horowitz Complexity Center Doctoral
fellowship. J.S. Langer was supported by U.S. Department of Energy
Grant No. DE-FG03-99ER45762. T.S. Lo was supported in part by the European Commission via a TMR grant.

\appendix

\section{Unbounded growth}
\label{cavitation}

One characteristic stress that emerges in the circular-hole problem
is the threshold $\sigma^{th}$ above which the hole grows without
bound under conditions of constant loading. In this section we
explain how this threshold is estimated in the athermal STZ theory.

Because the only length scale in this problem is the radius of the
hole, $R(t)$, we expect that all spatially varying ($r$-dependent)
quantities occurring in a uniformly (exponentially) expanding solution of our
equations of motion will be self-similar functions only of the ratio $r/R(t)$.
For convenience, we choose the variable $\xi=R(t)/r$.
These self-similar solutions are characterized by
$\dot{R}/R=\omega$, where $\omega>0$ depends on the loading and
vanishes at the threshold $\sigma^{th}$.

In fact, these solutions
invalidate the assumptions of our theory for large times since the
exponential growth of the hole implied by $\dot{R}/R=\omega$ will
inevitably lead to very large velocities at large $r$, in
contradiction with our
omission of the inertial term in Eq. (\ref{eqmot1}). When this
happens in the full, inertial theory, the self-similar solutions
break down with the appearance of perturbations propagating at
a characteristic wave velocity. Nevertheless, as we are interested
in obtaining an estimate for $\sigma^{th}$, the self-similar
approximate solutions for finite times are still useful so long
as we understand that we are actually dealing with a system in
which the remote tractions are being applied at a large but
finite distance from the hole.

For present purposes, we solve our equations of motion for a
self-similar deviatoric stress $\tilde s(\xi;\omega)$ in the
limit $\tau_0 \,\omega \to 0$. Once $\tilde s(\xi;\omega)$ is
known, we can use Eq. (\ref{integral}) to obtain the
following expression for $\sigma^{th}$
\begin{equation}
{\sigma^{th}\over s_y}\simeq 2\int_{0}^{1} \frac{\tilde s(\xi; \omega)}{\xi} d\xi
\quad \hbox{for}\quad \tau_0\,\omega\to 0 \ . \label{integral_SS}
\end{equation}
Here we assume that the exponential growth at a small $\omega$
allows the system to approach the self-similar
solution $\tilde s(\xi; \omega)$ in times when the velocities are
still smaller than the typical wave velocity. $\tilde s(\xi; \omega)$
can be computed using Eqs. (\ref{first_eq})-(\ref{STZ3_incom}).
First we note that the material time derivative of Eq.
(\ref{Mat_Der}) translates to
\begin{equation}
\frac{\cal D}{{\cal D} t} = \omega \,\xi\, (1-\xi^2)\,
\frac{\partial}{\partial
 \xi} \label{Mat_Der_xi} \ .
\end{equation}
Then Eqs. (\ref{first_eq})-(\ref{STZ3_incom}) read
\begin{eqnarray}
\tau_0\, \omega \,\xi \,(1-\xi^2) \,\frac{\partial \tilde s}{\partial
 \xi}&=& 2\,\mu\,\tau_0\,\omega\,\xi^2-2\,\mu\,\epsilon_0\, e^{-1/\chi}\,q(\tilde s, m;\zeta)
 \nonumber\\
\tau_0\, \omega\, \xi\, (1-\xi^2)\, \frac{\partial m}{\partial
 \xi}&=& 2\,q(\tilde s, m;\zeta)\,(1-\tilde s\,m)\nonumber\\
\tau_0 \,\omega\, \xi\, (1-\xi^2)\, \frac{\partial \chi}{\partial
 \xi}&=& {2\,\epsilon_0\over
c_0\,\tau_0}\, e^{-1/\chi}\,\tilde s\,q(\tilde s,m; \zeta)\,(\chi_{\infty}-\chi)
\ . \nonumber\\ \label{SS}
\end{eqnarray}
Also recall that $\dot{R}/R=\omega$. The initial conditions are
\begin{equation}
\tilde s(0)=0,\quad m(0)=0,\quad \chi(0)=\chi_0 \ .
\end{equation}
We have integrated these equations numerically in the limit
$\tau_0\,\!\omega\!\to 0$ with $\epsilon_0\!=\!1, \quad c_0\!=\!1,
\quad \chi_{\infty}\!=\!0.13, \quad \chi_0\!=\!0.1, \quad
\mu/s_y\!=\!50$ and for various values of $\zeta\!<\!10$. Then
we have used Eq. (\ref{integral_SS}) to obtain
$\sigma^{th} \approx 5\,s_y$, and have checked that this result is only
weakly dependent on $\zeta$.


\begin{thebibliography}{99}
\bibitem{Refs} References to other approaches to the circular hole problem can
be found in \cite{99LL}.
\end{thebibliography}
\end{document}